%% file: main.tex
\pdfoutput=1
\documentclass{article}

\usepackage{microtype}
\usepackage{graphicx}
\usepackage{subfigure}
\usepackage{booktabs} % for professional tables

\usepackage{hyperref}

\usepackage[accepted]{mlsys2021}

\mlsystitlerunning{QAPPA: Quantization-Aware Power, Performance, and Area Modeling of DNN Accelerators}

\begin{document}

\twocolumn[
\mlsystitle{QAPPA: Quantization-Aware Power, Performance, and Area Modeling of DNN Accelerators}

% unless you've provided the [accepted] option to the mlsys2021


% \mlsyssetsymbol{equal}{*}

% \begin{mlsysauthorlist}
% \mlsysauthor{Aeiau Zzzz}{equal,to}
% \mlsysauthor{Bauiu C.~Yyyy}{equal,to,goo}
% \mlsysauthor{Cieua Vvvvv}{goo}
% \mlsysauthor{Iaesut Saoeu}{ed}
% \mlsysauthor{Fiuea Rrrr}{to}
% \mlsysauthor{Tateu H.~Yasehe}{ed,to,goo}
% \mlsysauthor{Aaoeu Iasoh}{goo}
% \mlsysauthor{Buiui Eueu}{ed}
% \mlsysauthor{Aeuia Zzzz}{ed}
% \mlsysauthor{Bieea C.~Yyyy}{to,goo}
% \mlsysauthor{Teoau Xxxx}{ed}
% \mlsysauthor{Eee Pppp}{ed}
% \end{mlsysauthorlist}

\begin{mlsysauthorlist}
\mlsysauthor{Ahmet Inci}{to}
\mlsysauthor{Siri Garudanagiri Virupaksha}{to}
\mlsysauthor{Aman Jain}{to}
\mlsysauthor{Venkata Vivek Thallam}{to}
\mlsysauthor{Ruizhou Ding}{to}
\mlsysauthor{Diana Marculescu}{to,goo}
\end{mlsysauthorlist}

\mlsysaffiliation{to}{Department of Electrical and Computer Engineering, Carnegie Mellon University, Pittsburgh, Pennsylvania, USA}
\mlsysaffiliation{goo}{Department of Electrical and Computer Engineering, The University of Texas at Austin, Austin, Texas, USA}
% \mlsysaffiliation{ed}{School of Computation, University of Edenborrow, Edenborrow, United Kingdom}

\mlsyscorrespondingauthor{Ahmet Inci}{ainci@andrew.cmu.edu}
% \mlsyscorrespondingauthor{Eee Pppp}{ep@eden.co.uk}

\mlsyskeywords{Machine Learning, MLSys}

\vskip 0.3in

\begin{abstract}
As the machine learning and systems community strives to achieve higher energy-efficiency through custom DNN accelerators and model compression techniques, there is a need for a design space exploration framework that incorporates quantization-aware processing elements into the accelerator design space while having accurate and fast power, performance, and area models. In this work, we present \textit{QAPPA}, a highly parameterized quantization-aware power, performance, and area modeling framework for DNN accelerators. Our framework can facilitate the future research on design space exploration of DNN accelerators for various design choices such as bit precision, processing element type, scratchpad sizes of processing elements, global buffer size, device bandwidth, number of total processing elements in the the design, and DNN workloads. Our results show that different bit precisions and processing element types lead to significant differences in terms of performance per area and energy. Specifically, our proposed lightweight processing elements achieve up to $4.9 \times$ more performance per area and energy improvement when compared to INT16 based implementation. 
\end{abstract}
]



\printAffiliationsAndNotice{}  % leave blank if no need to mention equal contribution
% \printAffiliationsAndNotice{\mlsysEqualContribution} % otherwise use the standard text.

\section{Introduction}
\input{introduction.tex}
\label{submission}

% \textbf{Paper Deadline:}

% \medskip

\section{Related Work}
\input{relatedwork.tex}

\section{Methodology}
\input{methodology.tex}

\section{Results}

\input{results.tex}

\section{Conclusion}
\input{conclusion.tex}

% Note use of \abovespace and \belowspace to get reasonable spacing
% above and below tabular lines.

% \begin{table}[t]
% \caption{Classification accuracies for naive Bayes and flexible
% Bayes on various data sets.}
% \label{sample-table}
% \vskip 0.15in
% \begin{center}
% \begin{small}
% \begin{sc}
% \begin{tabular}{lcccr}
% \toprule
% Data set & Naive & Flexible & Better? \\
% \midrule
% Breast    & 95.9$\pm$ 0.2& 96.7$\pm$ 0.2& $\surd$ \\
% Cleveland & 83.3$\pm$ 0.6& 80.0$\pm$ 0.6& $\times$\\
% Glass2    & 61.9$\pm$ 1.4& 83.8$\pm$ 0.7& $\surd$ \\
% Credit    & 74.8$\pm$ 0.5& 78.3$\pm$ 0.6&         \\
% Horse     & 73.3$\pm$ 0.9& 69.7$\pm$ 1.0& $\times$\\
% Meta      & 67.1$\pm$ 0.6& 76.5$\pm$ 0.5& $\surd$ \\
% Pima      & 75.1$\pm$ 0.6& 73.9$\pm$ 0.5&         \\
% Vehicle   & 44.9$\pm$ 0.6& 61.5$\pm$ 0.4& $\surd$ \\
% \bottomrule
% \end{tabular}
% \end{sc}
% \end{small}
% \end{center}
% \vskip -0.1in
% \end{table}

\section*{Acknowledgements}

This research was supported in part by National Science Foundation grants CCF No. 1815899 and CSR No. 1815780.


\bibliography{main}
\bibliographystyle{mlsys2021}

%%%%%%%%%%%%%%%%%%%%%%%%%%%%%%%%%%%%%%%%%%%%%%%%%%%%%%%%%%%%%%%%%%%%%%%%%%%%%%%
%%%%%%%%%%%%%%%%%%%%%%%%%%%%%%%%%%%%%%%%%%%%%%%%%%%%%%%%%%%%%%%%%%%%%%%%%%%%%%%
% SUPPLEMENTAL CONTENT AS APPENDIX AFTER REFERENCES
%%%%%%%%%%%%%%%%%%%%%%%%%%%%%%%%%%%%%%%%%%%%%%%%%%%%%%%%%%%%%%%%%%%%%%%%%%%%%%%
%%%%%%%%%%%%%%%%%%%%%%%%%%%%%%%%%%%%%%%%%%%%%%%%%%%%%%%%%%%%%%%%%%%%%%%%%%%%%%%
% \appendix
% \section{Please add supplemental material as appendix here}
% %
% Put anything that you might normally include after the references as an appendix here, {\it not in a separate supplementary file}. Upload your final camera-ready as a single pdf, including all appendices.

%%%%%%%%%%%%%%%%%%%%%%%%%%%%%%%%%%%%%%%%%%%%%%%%%%%%%%%%%%%%%%%%%%%%%%%%%%%%%%%
%%%%%%%%%%%%%%%%%%%%%%%%%%%%%%%%%%%%%%%%%%%%%%%%%%%%%%%%%%%%%%%%%%%%%%%%%%%%%%%

\end{document}

%% file: introduction.tex
Deep neural networks (DNNs) have achieved remarkable accomplishments across various applications ranging from image recognition \cite{EfficientNet}, object detection \cite{EfficientDet}, to natural language processing \cite{Devlin2019BERTPO}. However, the increasing model size and computational cost of these models become a challenging task for on-device machine learning (ML) endeavours due to the stringent performance per area and energy constraints of the edge devices. To this end, while machine learning practitioners focus on model compression techniques \cite{han2016deep,ruizhou2018lightnn,Chin2020CVPR}, computer architects investigate hardware architectures to overcome the energy-efficiency problem and improve the overall system performance \cite{inci2018asbd,DeepNVM,inci2021tcad,inci2020architectural,inci2021cross}.

As computing community hits the limits on consistent performance scaling for traditional architectures, there has been a rising interest on enabling on-device machine learning through custom DNN accelerators. As we deeply care about performance per area and energy-efficiency from a hardware point of view, tailored DNN accelerators have shown significant improvements when compared to CPUs and GPUs \cite{eyeriss,tpu,tetris}. To better understand the trade-offs of various architectural design choices and DNN workloads, there is a need for a design space exploration framework that can rapidly iterate over various designs and generate power, performance, and area (PPA) results. To this end, in this work we present \textit{QAPPA}, a quantization-aware power, performance, and area modeling framework for DNN accelerators.

\begin{figure}[t]
  \centering
 \includegraphics[width=0.5\textwidth]{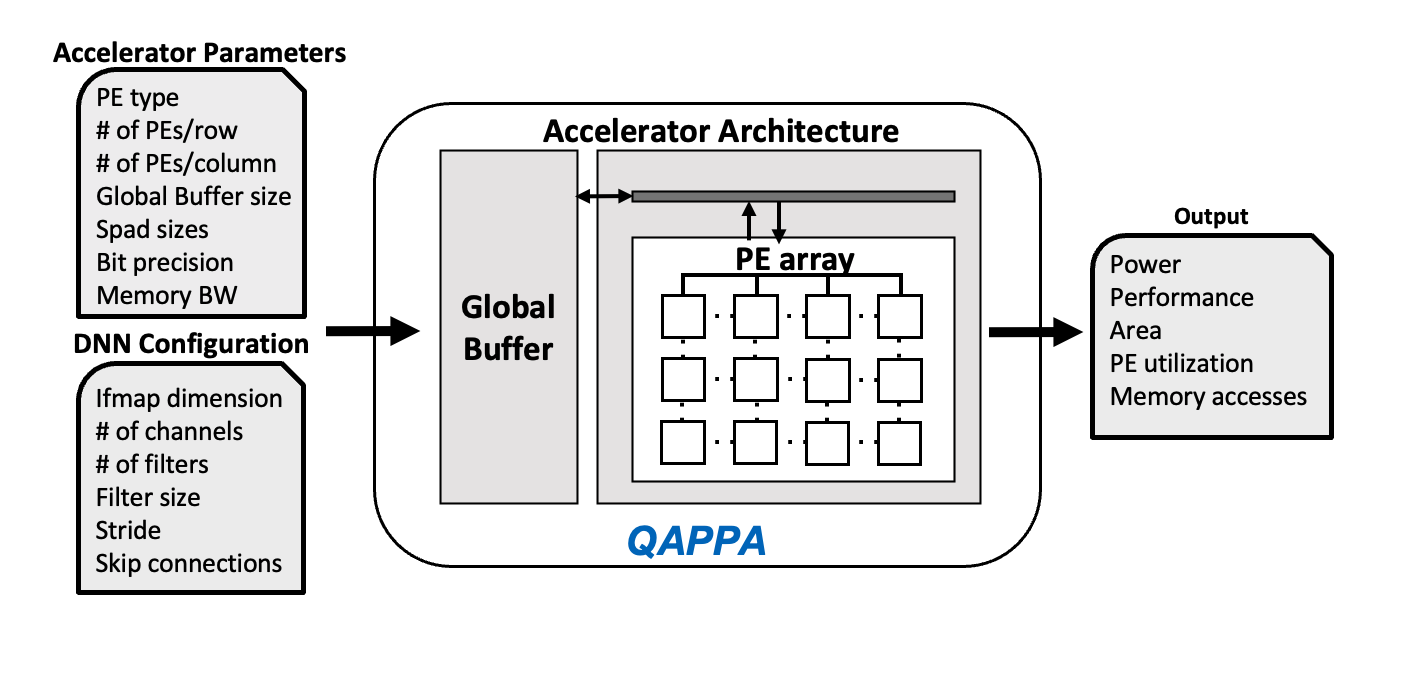}
 \caption{{Schematic depicting \textit{QAPPA} framework, with accelerator parameters and DNN configuration as inputs. The framework takes in accelerator parameters and layer-wise DNN configurations and generates power, performance, area results, and statistics on hardware utilization and memory accesses.}}\label{fig:qappa}
\end{figure}

This work makes the following contributions: 
\begin{itemize}
    \item We present QAPPA, a quantization-aware power, performance, and area modeling framework for DNN accelerators. Our framework can enable future research on design space exploration of DNN accelerators for various design choices such as bit precision, processing element types, scratchpad size of processing elements, global buffer size, device bandwidth, number of total processing elements in the the design, and DNN workloads. 
    \item  Our framework provides power, performance, and area results not just for a single hardware design point but for a range of different hardware designs as opposed to prior art \cite{qi17paleo,cai2017neuralpower}. Therefore, it can be used to analyze trade-offs of various architectural design choices and DNN workloads at the same time.  
    % \item 
    
\end{itemize}

The rest of the paper is organized as follows. In Section 2, we present a literature review on power and runtime models for CNNs and design space exploration frameworks for hardware accelerators. 
In Section 3, we describe the architectural details of the \textit{QAPPA} framework and the details of our methodology for power, performance, and area modeling of DNN accelerators. In Section 4, we show experimental results demonstrating the efficiency of \textit{QAPPA}'s power, performance, area models and the efficacy of lightweight processing elements to conventional designs in terms of performance per area and energy through a suite of case studies. 
Finally, Section 5 concludes the paper by summarizing the results.

%% file: relatedwork.tex
Prior art has proposed runtime and energy models for DNN workloads \cite{cai2017neuralpower,qi17paleo}. However, these models have been implemented specifically for GPU platforms and thus they create an important limitation for a design space exploration of hardware architectures and potentially hardware/ML model co-design opportunities \cite{gupta2020acceleratoraware,codesign_vikas}. On the other hand, prior art has proposed tools and simulation methodologies for accelerator design. For example, SCALE-Sim \cite{samajdar2018scale} is a cycle accurate, systolic-array based DNN accelerator simulator. Similarly, Aladdin \cite{Shao2014AladdinAP} is a pre-RTL power and performance accelerator simulator. Although these tools help to perform preliminary analysis on the design space for accelerators in different aspects, they do not incorporate quantization-aware processing elements and they do not generate RTL output based on the input hardware configuration which is an important impediment for enabling deployment of DNNs onto edge devices.

%% file: methodology.tex
In this section, we first explain the implementation details and architectural components of our \textit{QAPPA} framework, as depicted in Figure~\ref{fig:qappa}. Next, we detail the lightweight processing elements (LightPE) that we implemented in our framework to provide a specialized processing element (PE) type for quantized DNN models. Finally, we explain our power, performance, and area modeling and design space exploration methodology.

\begin{figure*}[t]
  \centering
 \includegraphics[width=0.24\textwidth]{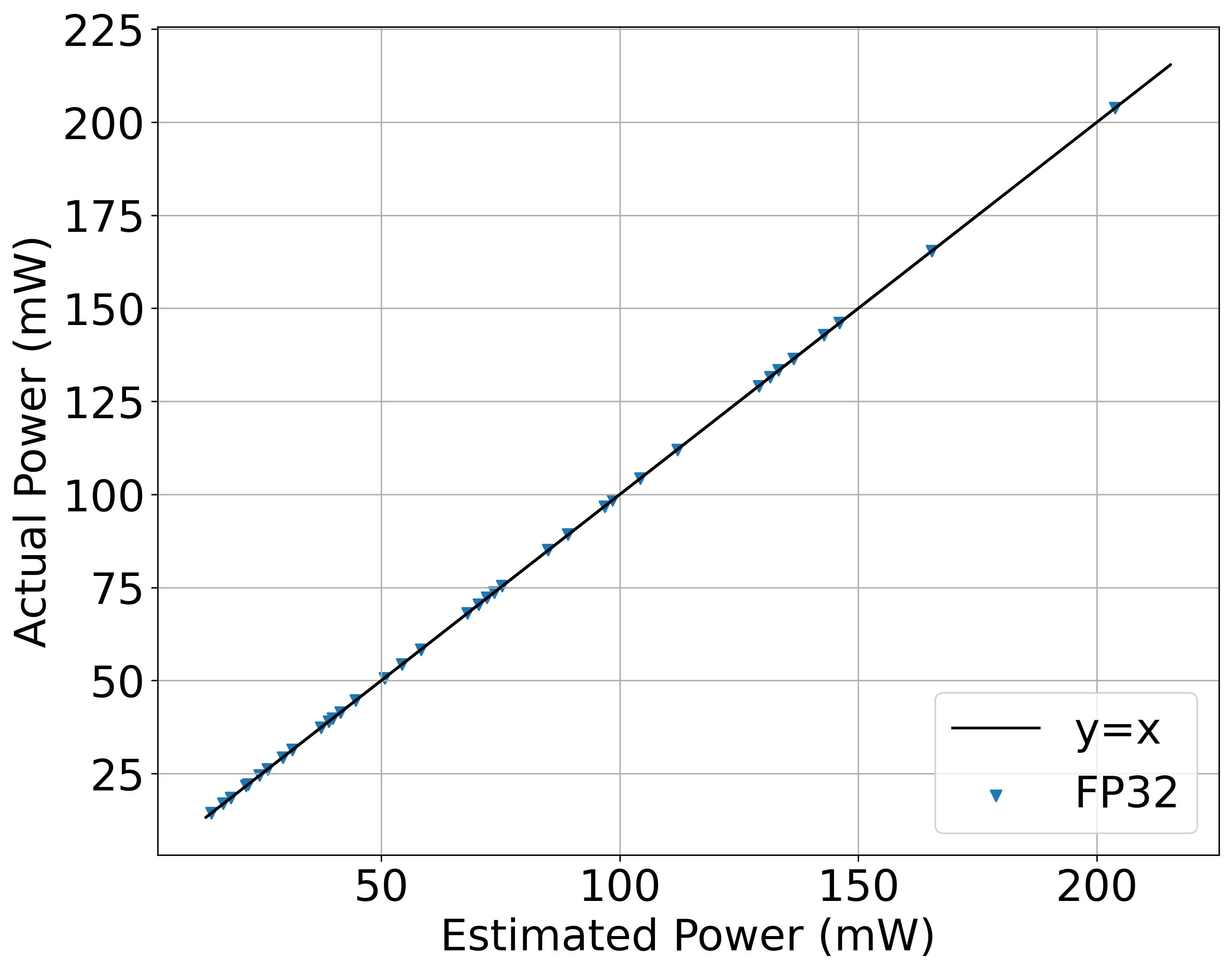}
  \includegraphics[width=0.24\textwidth]{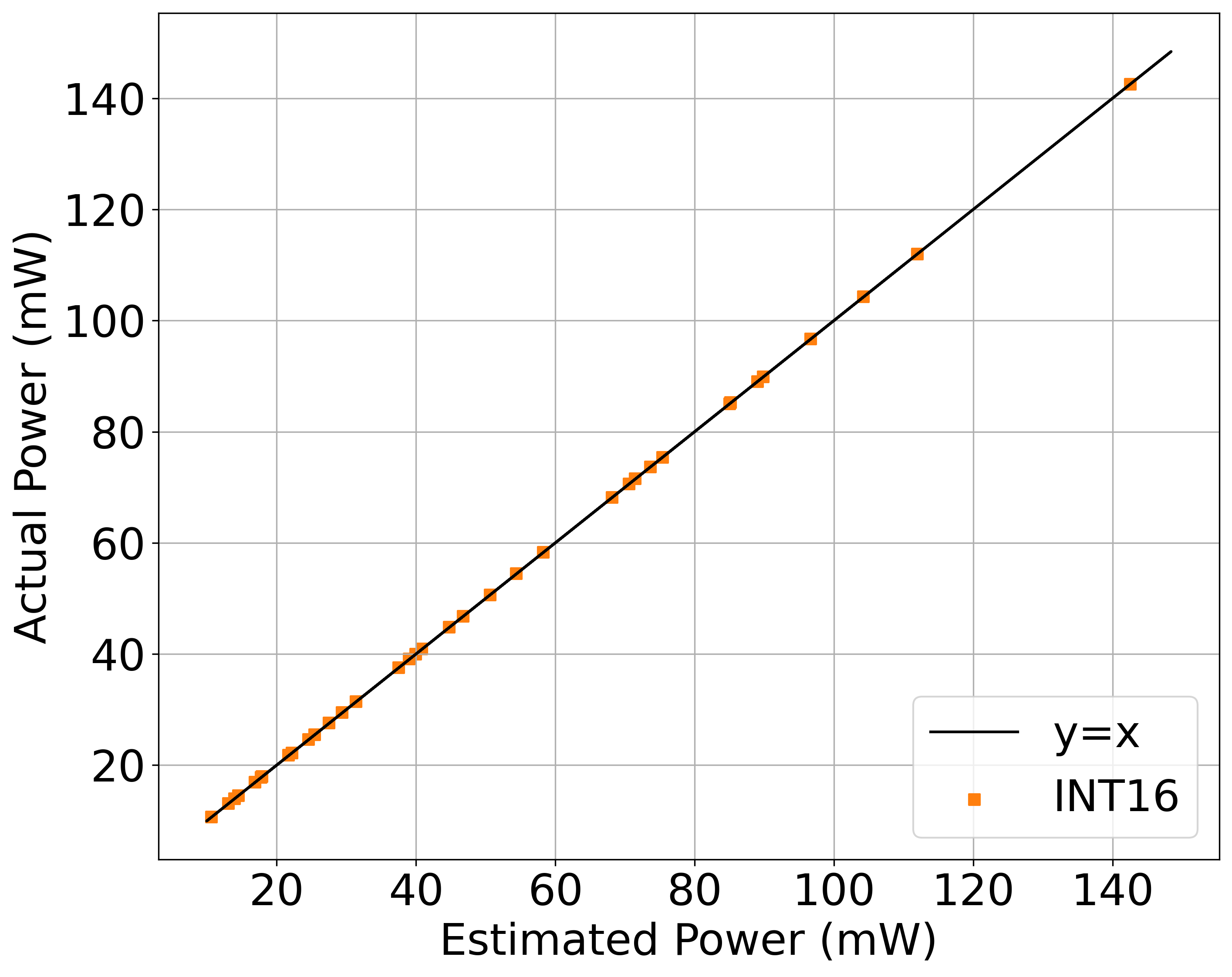}
  \includegraphics[width=0.24\textwidth]{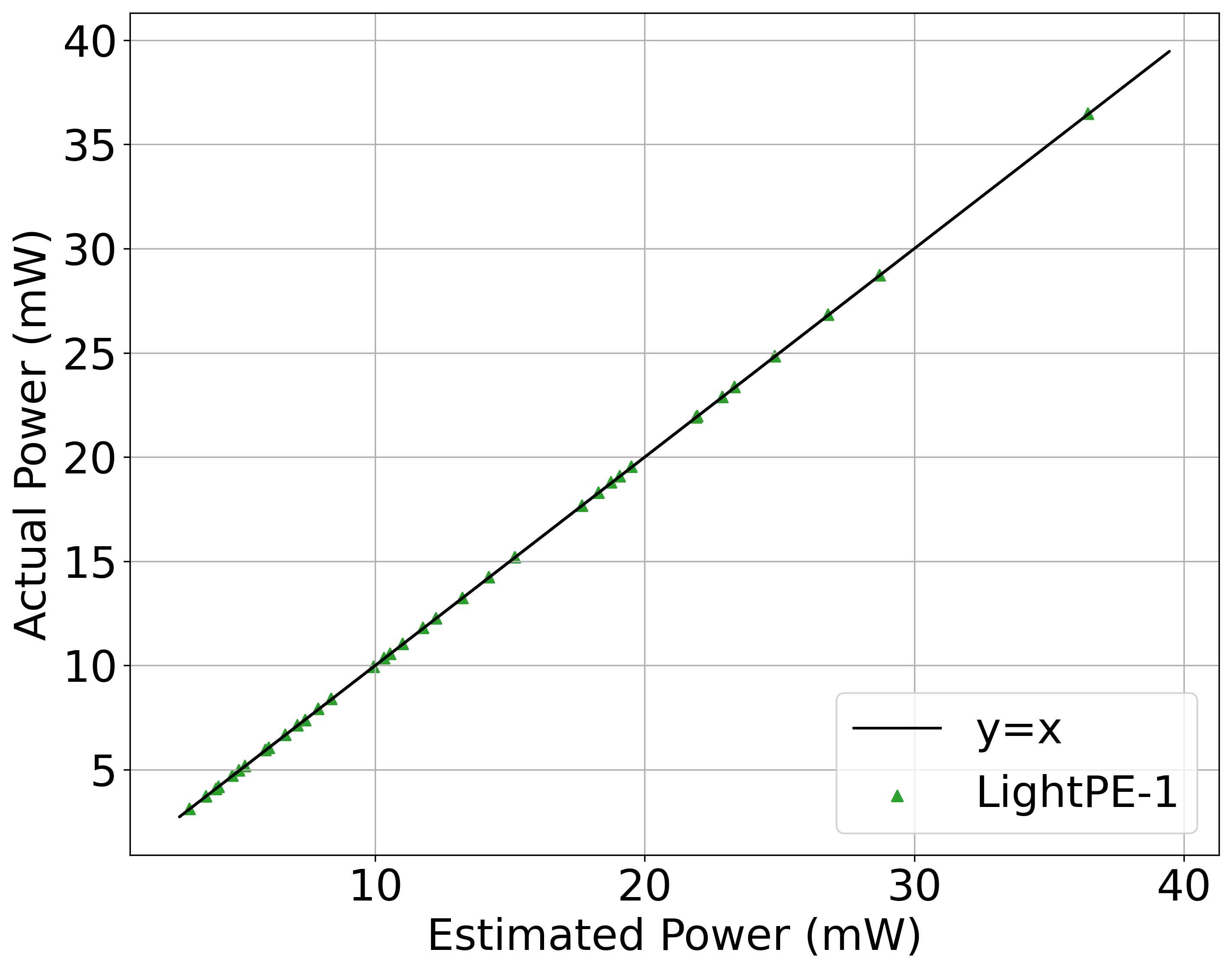}
    \includegraphics[width=0.24\textwidth]{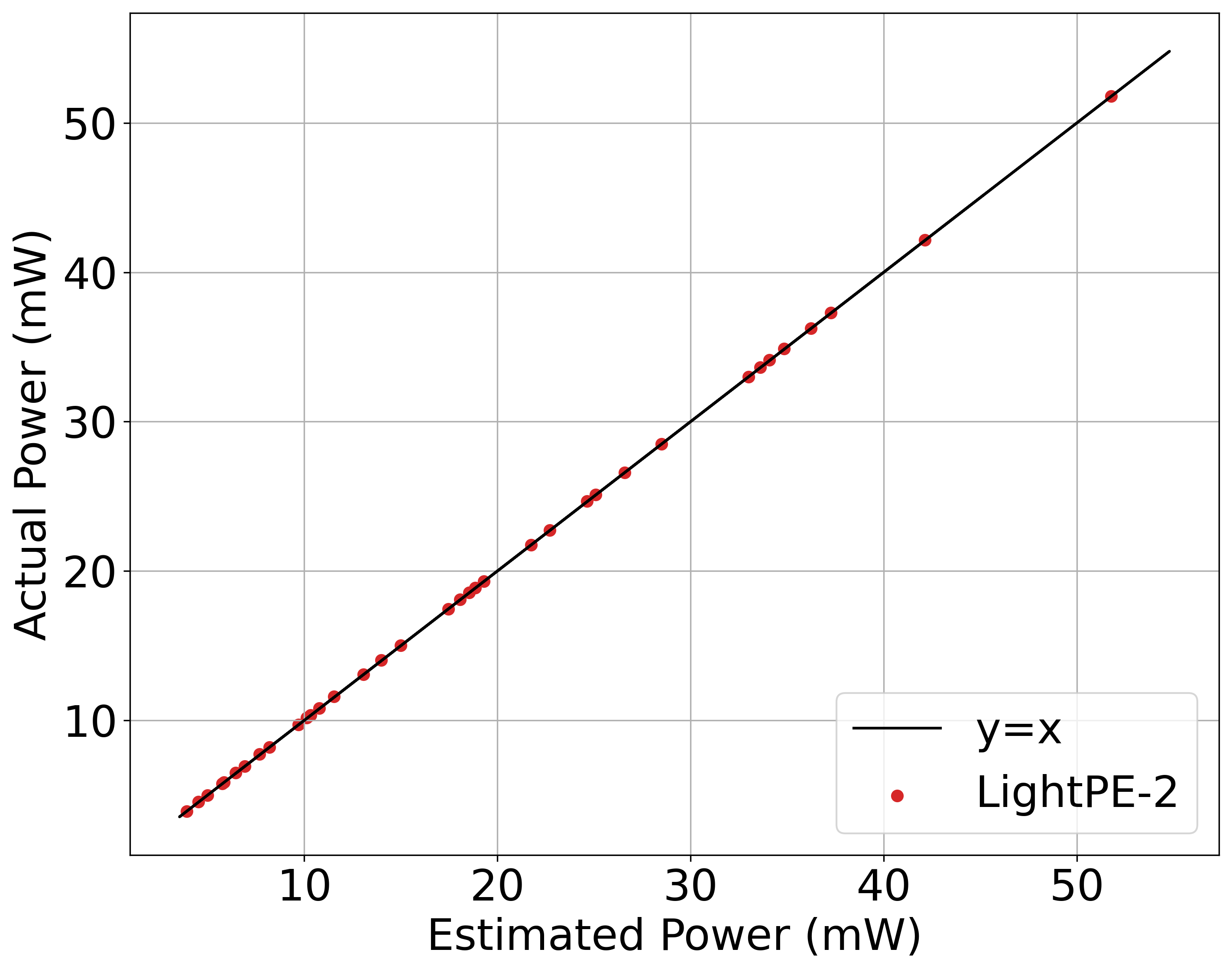}
    \\
     \includegraphics[width=0.24\textwidth]{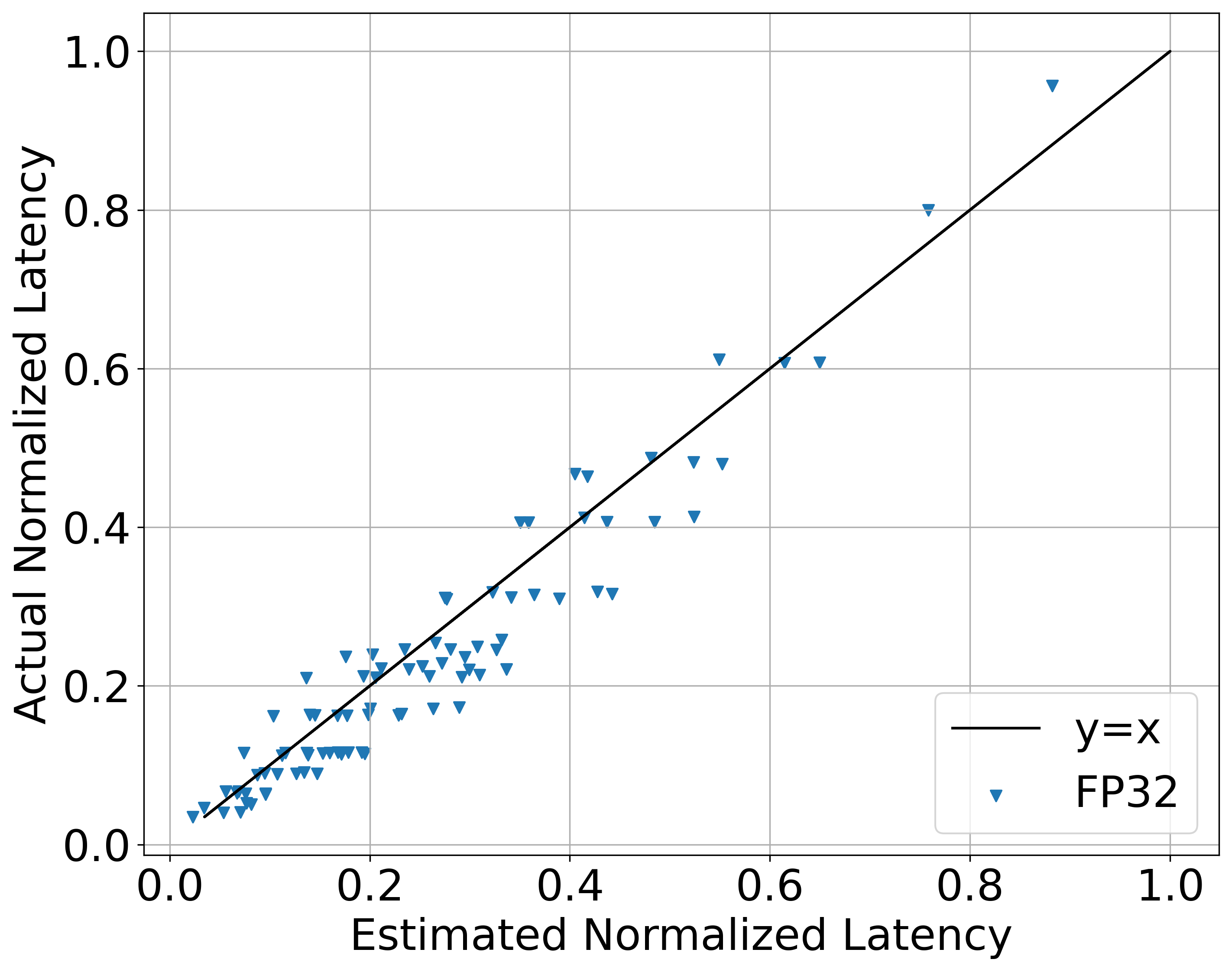}
  \includegraphics[width=0.24\textwidth]{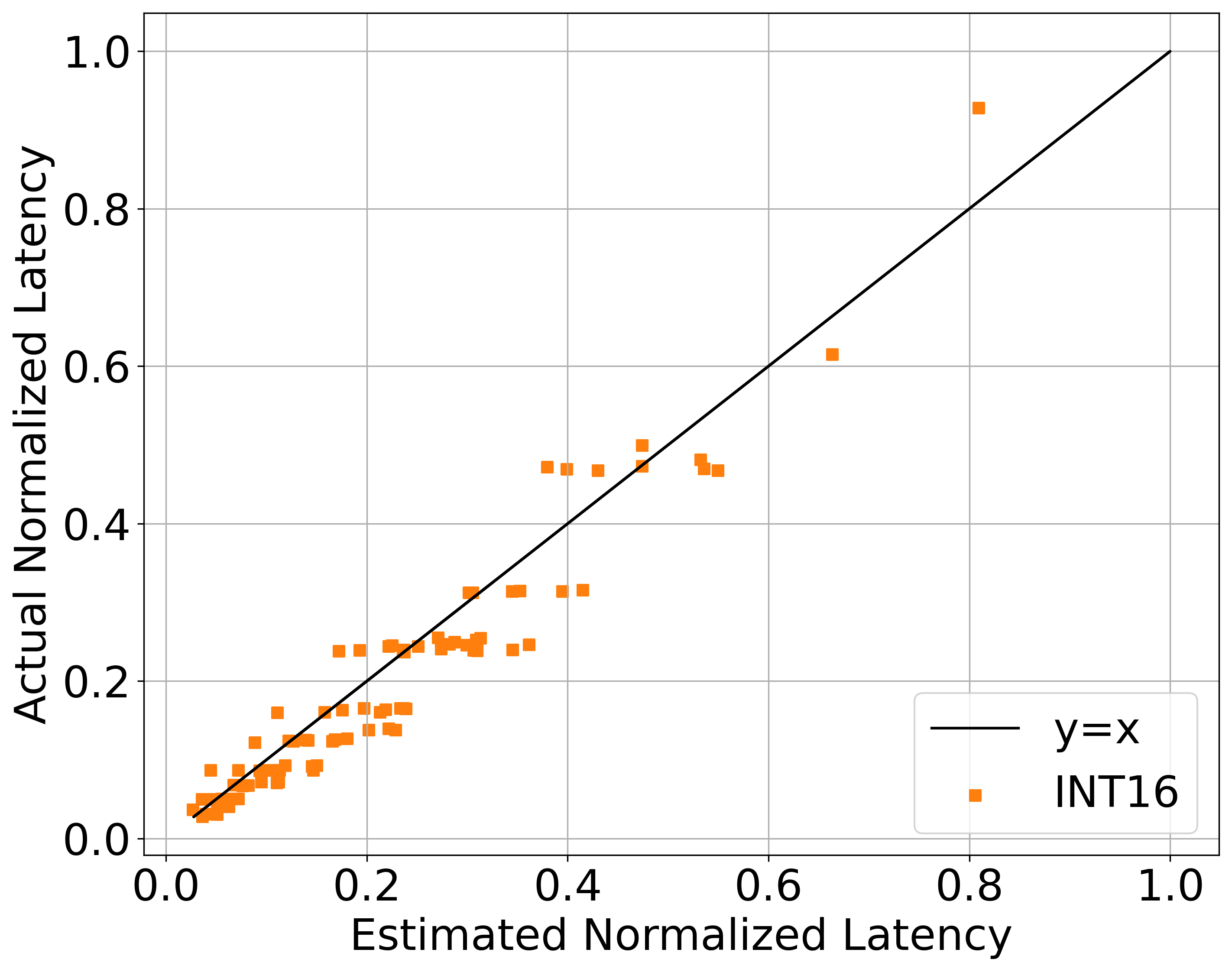}
  \includegraphics[width=0.24\textwidth]{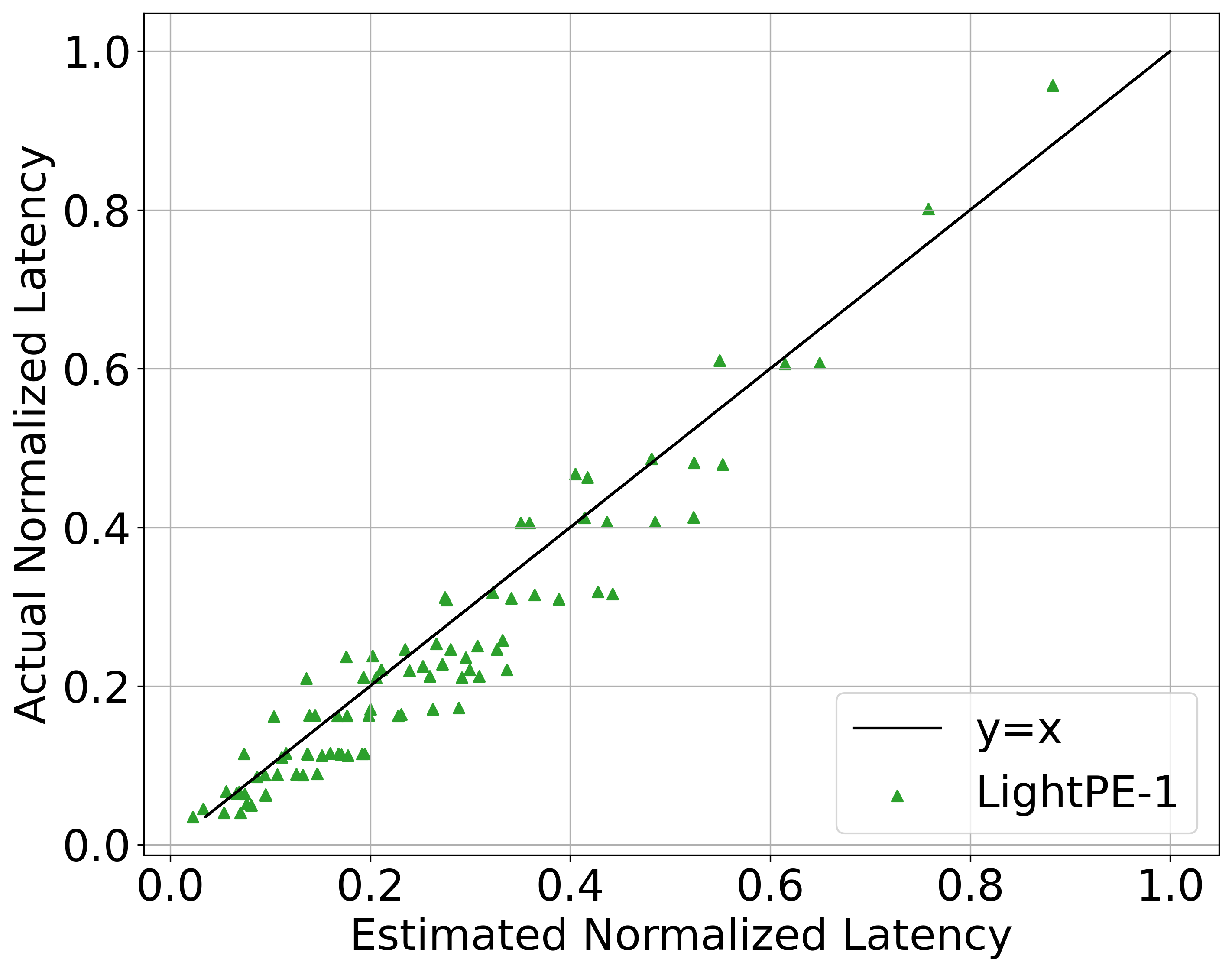}
    \includegraphics[width=0.24\textwidth]{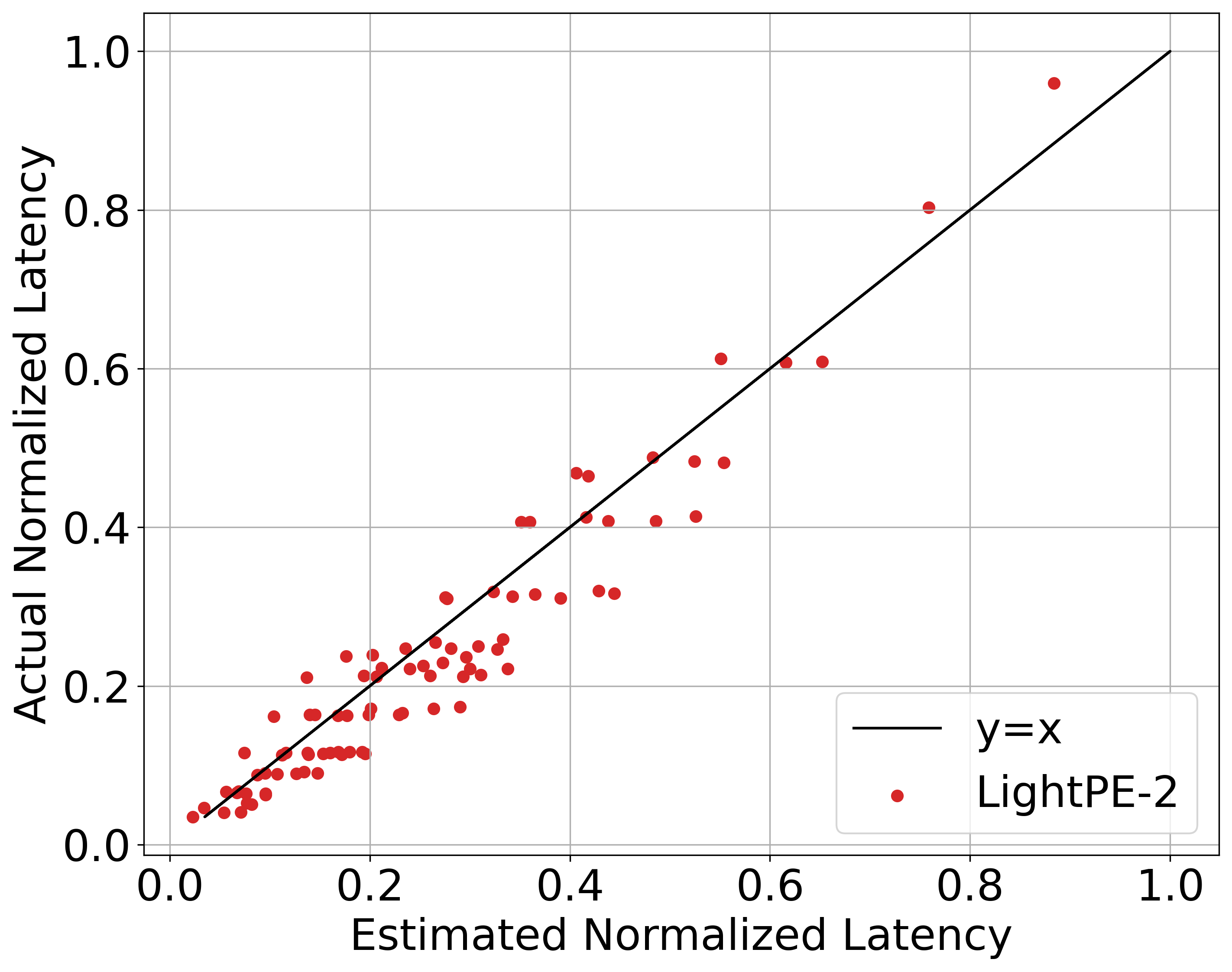}
    \\
 \includegraphics[width=0.24\textwidth]{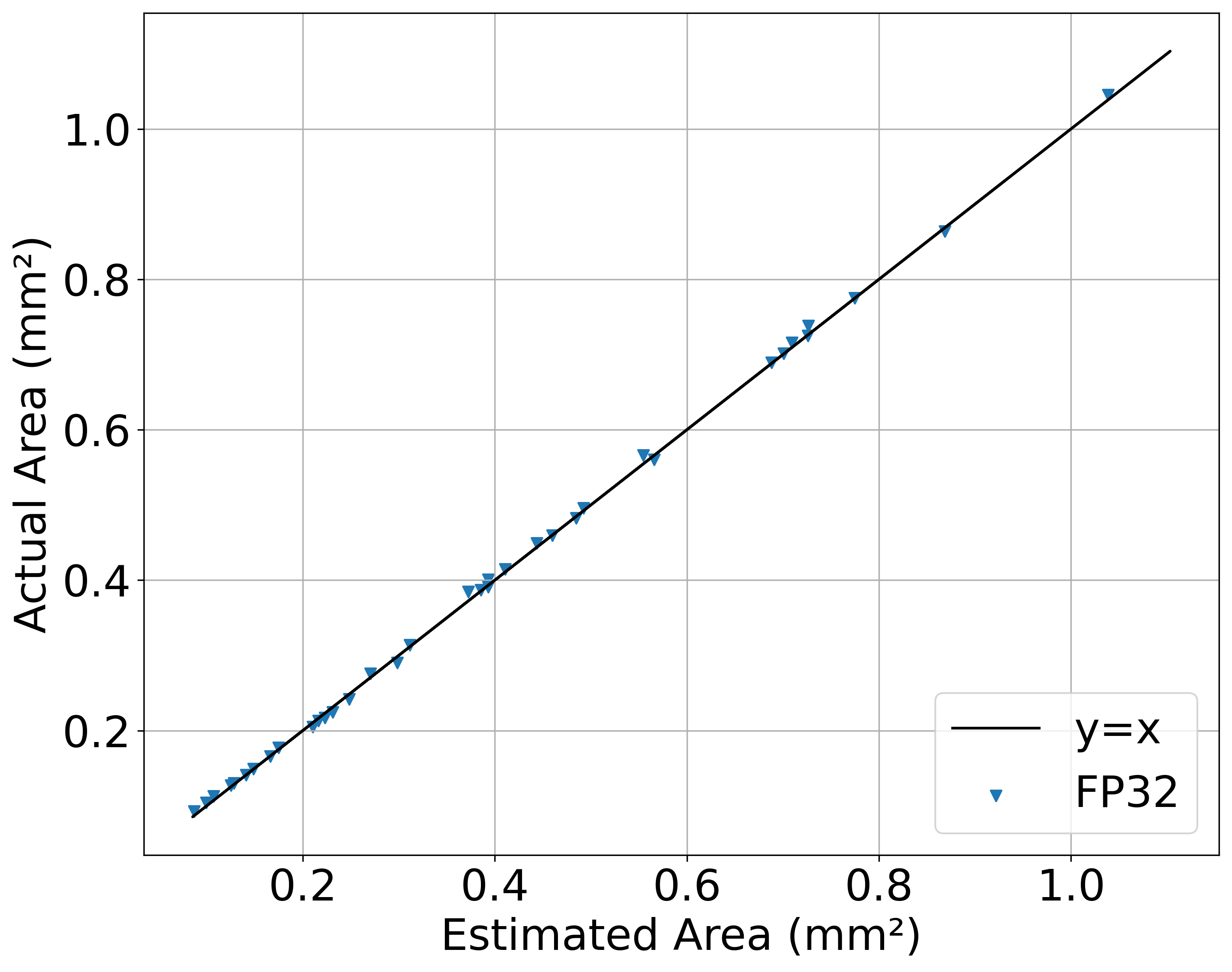}
  \includegraphics[width=0.24\textwidth]{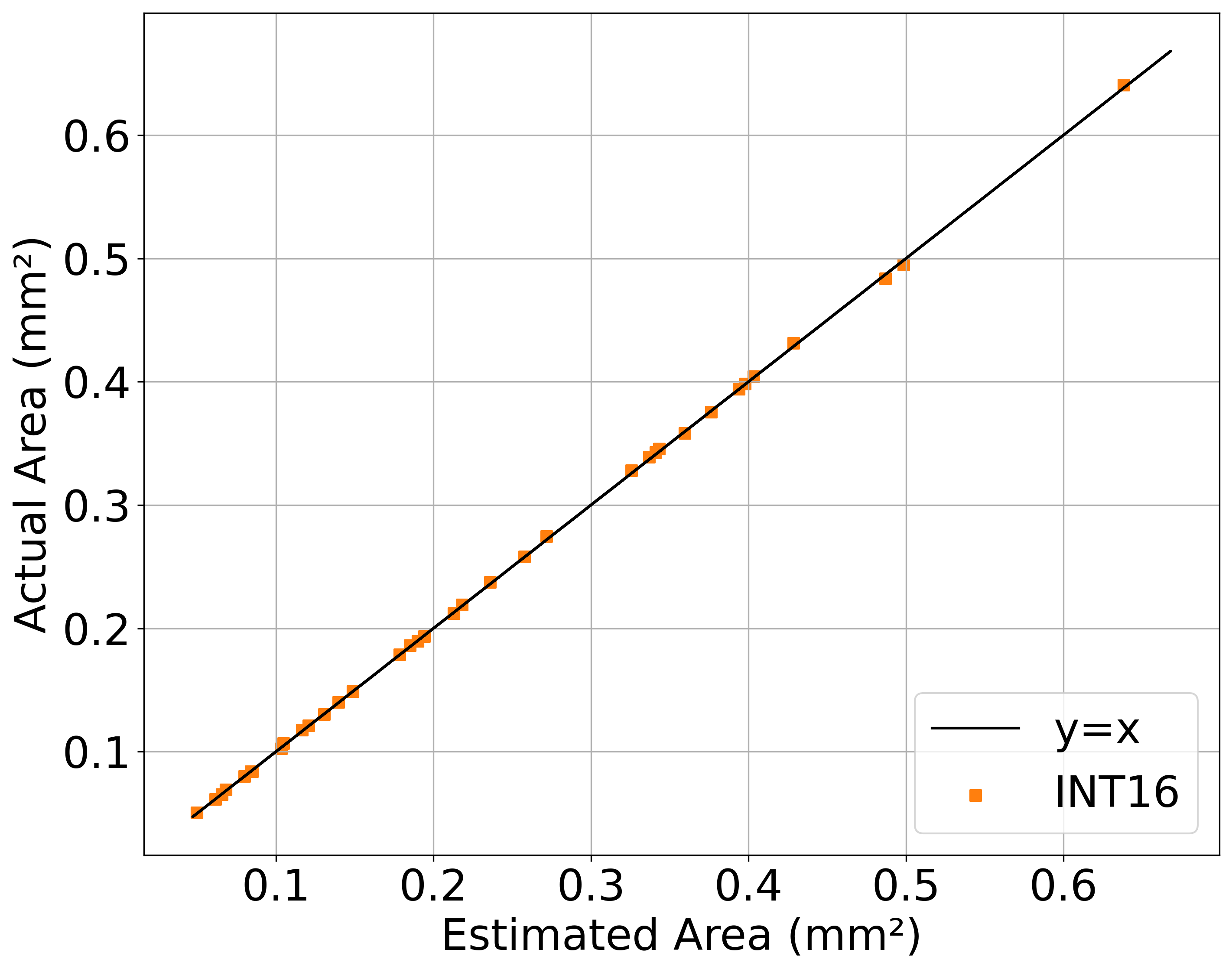}
  \includegraphics[width=0.24\textwidth]{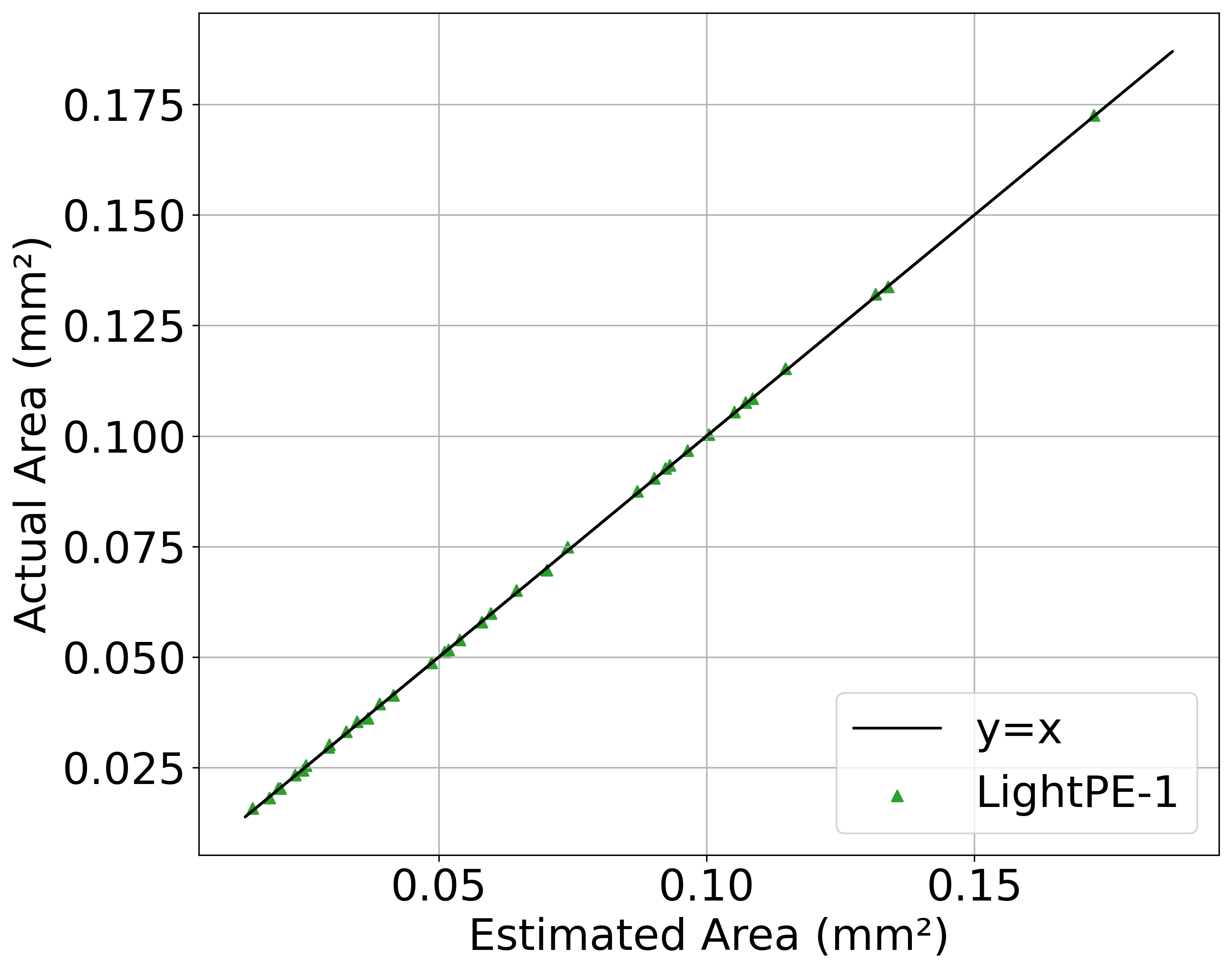}
    \includegraphics[width=0.24\textwidth]{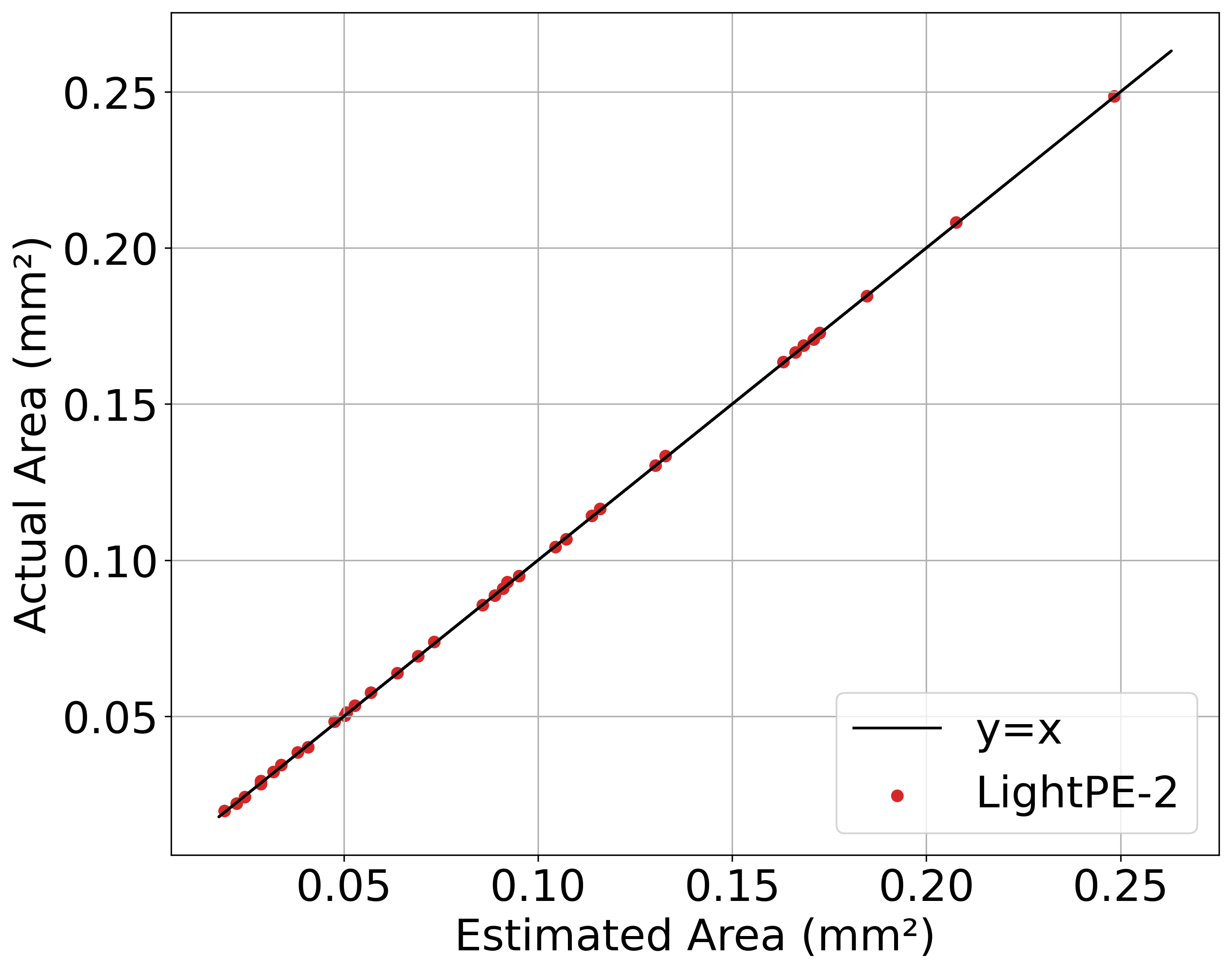}
    
 \caption{{Power (top chart), performance (middle chart), and area (bottom chart) estimation results for various processing element types such as FP32, INT16, LightPE-1, and LightPE-2. Each data point corresponds to a different hardware configuration that can be achieved by using the corresponding processing element type. As it can be seen, the proposed polynomial model agrees closely with the actual values extracted from the synthesis tools.}}\label{fig:model}
\end{figure*}

\subsection{QAPPA Framework}

To enable comprehensive design space exploration for DNN accelerators for on-device machine learning, we implemented \textit{QAPPA}, a highly parameterized spatial-array based DNN accelerator framework in RTL. Our framework enables hardware designers and machine learning practitioners to rapidly iterate over various accelerator designs and DNN configurations and better understand trade-offs of different architectural components of the design for dizzying requirements of deploying machine learning models to edge devices. Moreover, hardware designers can also use the automatically generated RTL code to follow the design synthesis flow. 

As depicted in Figure~\ref{fig:qappa}, \textit{QAPPA} framework is based on spatial-array based accelerators and utilizes row stationary dataflow which has been demonstrated to optimize the data movement in the storage hierarchy \cite{eyeriss}. \textit{QAPPA} features a set of processing elements organized as a 2D array and a global buffer that stores input feature maps, filters, and activations. The number of PEs in each dimension can be tuned for different power, performance, and area requirements. In each PE, there are input feature map, filter, and partial sum scratchpads and a multiply-accumulate (MAC) unit which can be changed based on the desired bit precision. Each of these architectural components can be tuned in a flexible and automated manner to perform a comprehensive design space exploration for on-device edge accelerators. 

\textbf{Lightweight Processing Elements (LightPE)} 

To enrich the design space of hardware accelerators and achieve a better Pareto-frontier in terms of performance per area and energy-efficiency perspectives, we include LightPE implementations in our framework. LightPEs utilize 8 bits for activations and 4 bits and 8 bits for weights for LightPE-1 and LightPE-2 designs, respectively. As 4 bit and 8 bit quantization techniques for on-device machine learning became prevalent in various computing platforms, we provide these specialized quantization-aware PE types in our \textit{QAPPA} framework to help hardware designers to enrich their design space and hopefully find better Pareto-frontiers.

Besides their low-precision benefits such as reducing the storage requirements, LightPEs also replace the multiplications with more energy and area-efficient one shift or a limited number of shifts and add operations \cite{ruizhou2018lightnn}. Therefore, they also achieve significant power and area gains when compared to full-precision 32 bit floating point and 16 bit integer based designs with only slight accuracy degradation \cite{ruizhou2018lightnn}. As a results, LightPEs provide an enriched design space for hardware designers to analyze various trade-offs between performance per area and energy.

\textbf{Power, Performance, and Area Modeling} 

To build our quantization-aware power, performance, and area models, we use various hardware and DNN configurations. Specifically, to cover this comprehensive design space of hardware accelerators, we run experiments by varying global buffer size, number of PEs per row and column in the 2D PE array, bit precision, and PE type (FP32, INT16, LightPE-1, and LightPE-2). Within each PE, we also vary individual scratchpad sizes for input feature map, filter scratch pad, and partial sum scratchpad. 

We use Synopsys Design Compiler and the open-source FreePDK45 which is a commonly used process design kit \cite{freepdk} to synthesize our designs to obtain power, area, and initial timing results. We use Synopsys VCS RTL simulator to perform functional verification and collect timing information for various DNN configurations such as VGG-16 \cite{vgg16}, ResNet-34, and ResNet-50 \cite{resnet50} that are implemented in our testbenches. 
After collecting power, area, and timing results from these tools, we use polynomial regression models and model selection techniques based on \textit{k}-fold cross validation \cite{kfold} to tune the model parameters and fit the model.

%% file: results.tex
In this section, we present power, performance, and area modeling results for each processing element type and perform a design space exploration on VGG-16 \cite{vgg16}, ResNet-34, and ResNet-50 \cite{resnet50} design spaces to iterate through our framework to demonstrate the flexibility of \textit{QAPPA} for future studies. 

As detailed in Section 3, \textit{QAPPA} framework provides power, performance, and area models that significantly speed up the design space exploration. Figure~\ref{fig:model} shows the actual and estimated power, performance, and area results for each processing element type such as FP32, INT16, LightPE-1, and LightPE-2. Each data point in Figure~\ref{fig:model} corresponds to a different hardware accelerator configuration in the comprehensive design space. As shown by the results, \textit{QAPPA}'s PPA models achieve high correlation to the actual PPA values. 
Figure~\ref{fig:model} also shows that the FP32 implementation has the highest area and power cost whereas LightPEs have the lowest area and power results when one processing element is considered. This shows the hardware-efficiency of LightPEs when compared to conventional PE implementations.

\begin{figure}[t]
  \centering
 \includegraphics[width=0.5\textwidth]{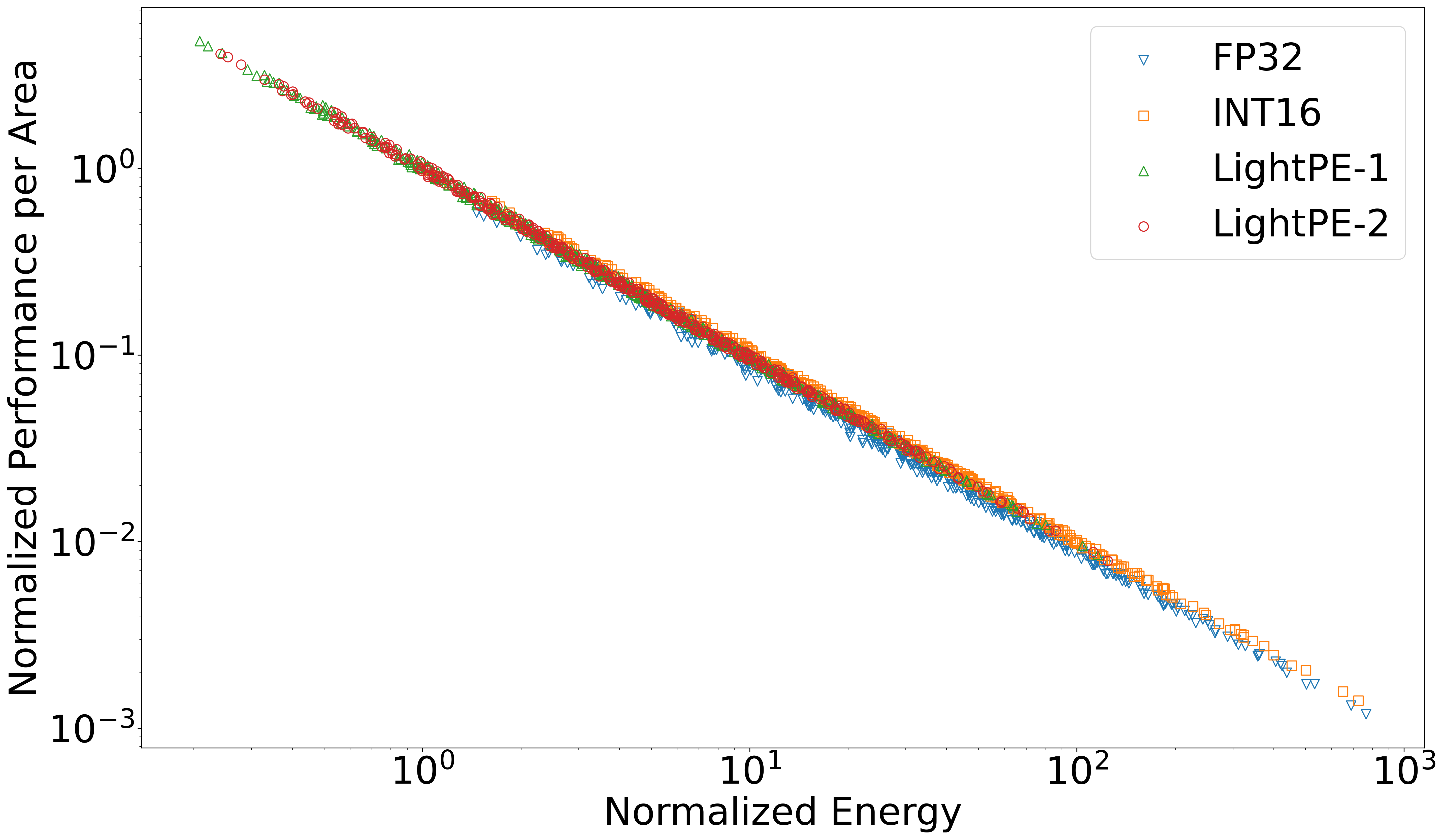}
 \caption{{Normalized performance per area vs. normalized energy results with respect to the INT16 hardware configuration with the highest performance per area for VGG-16 design space}}\label{fig:vgg}
\end{figure}

\begin{figure}[t]
  \centering
 \includegraphics[width=0.5\textwidth]{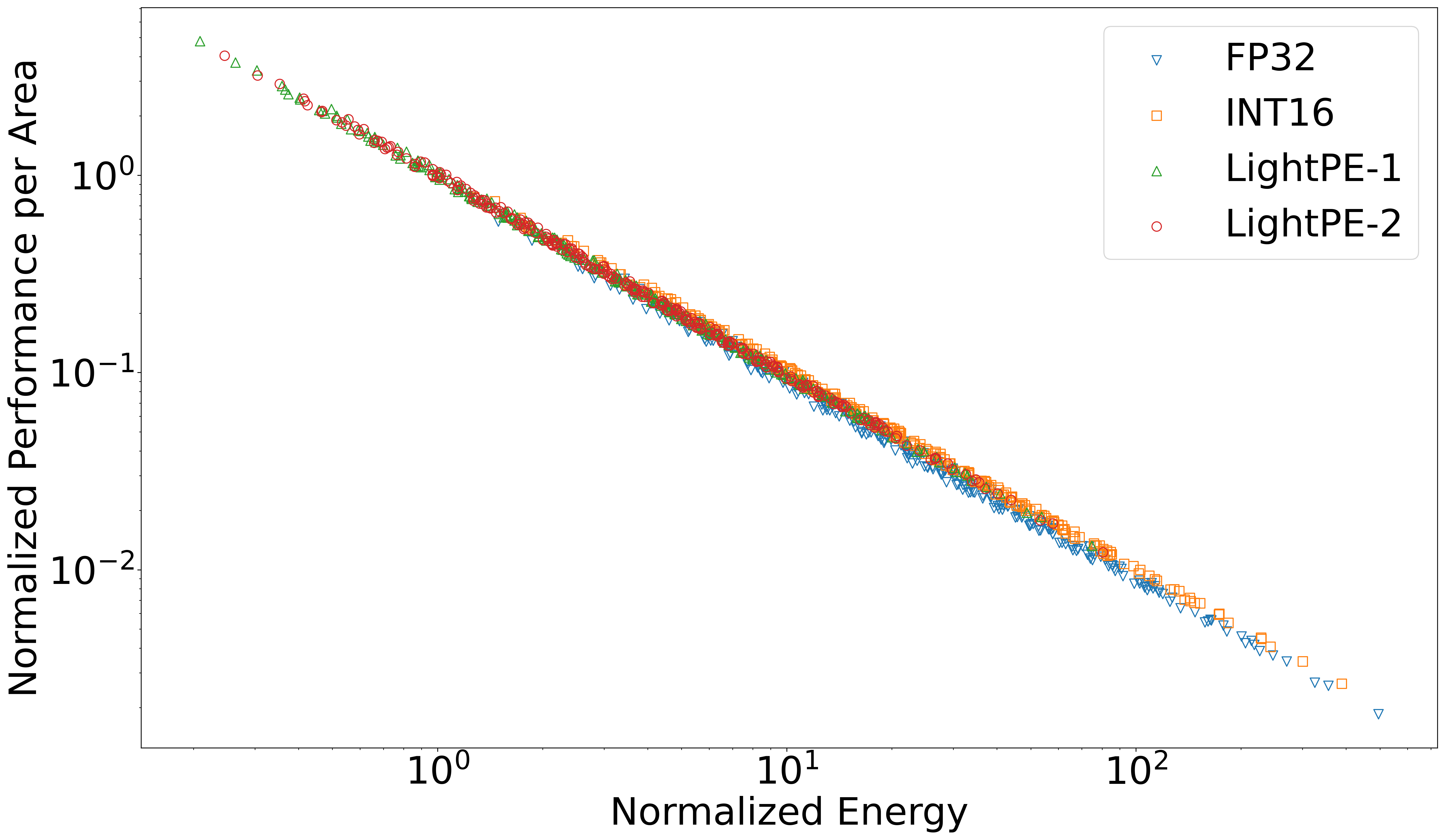}
 \caption{{Normalized performance per area vs. normalized energy results with respect to the INT16 hardware configuration with the highest performance per area for ResNet-34 design space}}\label{fig:res34}
\end{figure}

\begin{figure}[t]
  \centering
 \includegraphics[width=0.5\textwidth]{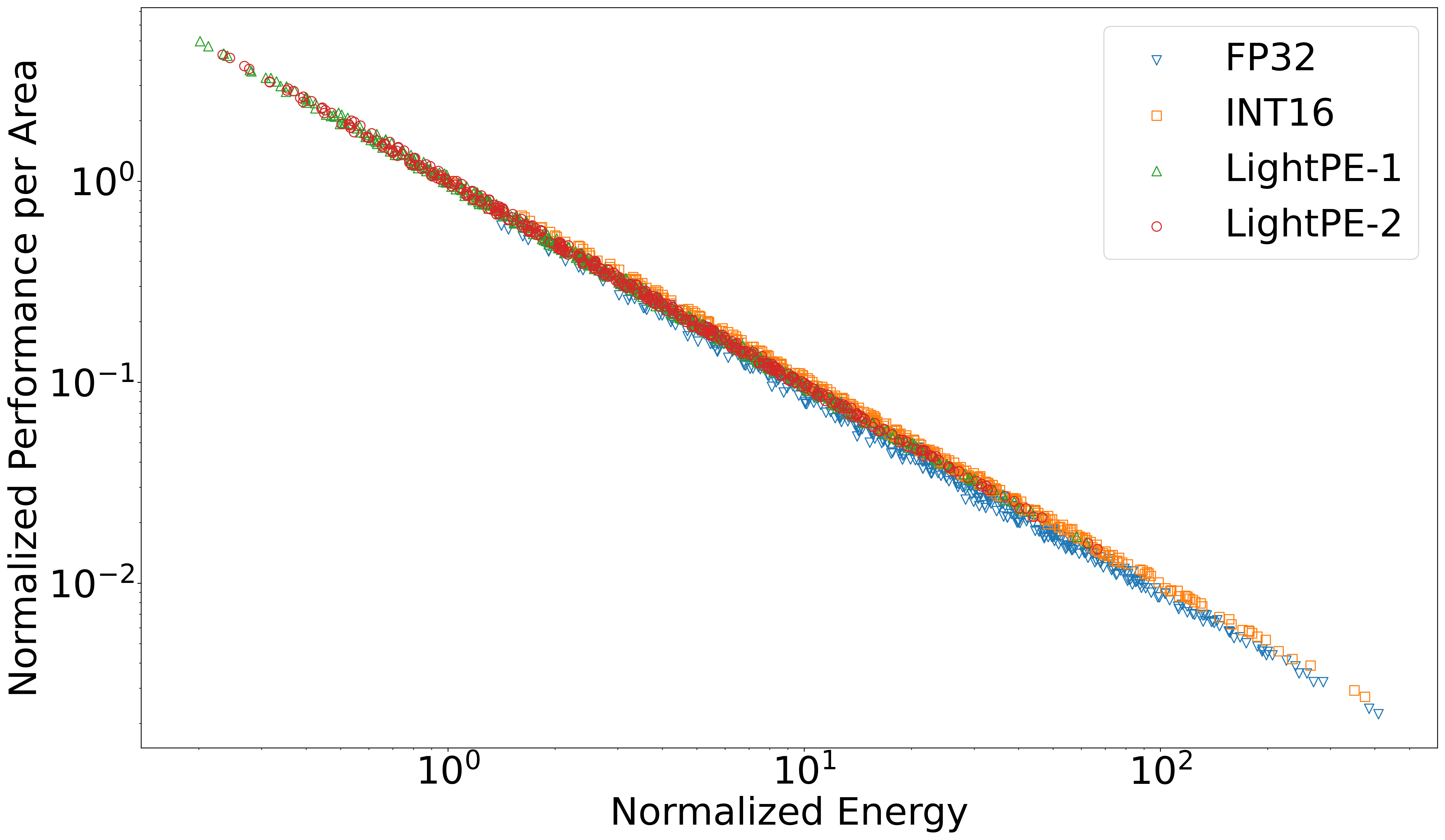}
 \caption{{Normalized performance per area vs. normalized energy results with respect to the INT16 hardware configuration with the highest performance per area for ResNet-50 design space}}\label{fig:res50}
\end{figure}

To show the efficacy of LightPEs to conventional PE designs, we perform design space exploration on  VGG-16 \cite{vgg16}, ResNet-34, and ResNet-50 \cite{resnet50} design spaces as shown in Figure~\ref{fig:vgg}-\ref{fig:res50}. 
We show the normalized performance per area and normalized energy results for each PE type with respect to the baseline INT16 based implementation with the highest performance per area for the given design space.

Figure~\ref{fig:vgg}-\ref{fig:res50} shows that LightPE implementations consistently outperform conventional INT16 and FP32 implementations in both aspects, which proves their efficacy in terms of hardware-efficiency. Specifically, LightPE-1 and LightPE-2 achieve $4.9 \times$ and $4.1 \times$ more performance per area and $4.9 \times$ and $4.2 \times$ energy improvement on average when compared to the best INT16 hardware configuration, respectively. On the other hand, INT16 baseline implementation achieves  $1.7 \times$ more performance per area and $1.4 \times$ energy improvement on average when compared to the best FP32 configuration. 
These conclusions hold for all models considered in this work VGG-16, ResNet-34, and ResNet-50, thereby showing that the benefits of using lower precision generalize across a variety of models. 
We conclude that different bit precisions and PE types can lead to significantly different performance per area and energy results which are two critical metrics for hardware designers and machine learning practitioners strive to improve upon. 

% \begin{figure*}[t]
%   \centering
%  \includegraphics[width=0.33\textwidth]{figures/clk_vgg2.png}
%   \includegraphics[width=0.33\textwidth]{figures/clk_vgg2.png}
%  \includegraphics[width=0.33\textwidth]{figures/clk_vgg2.png}

%  \caption{{Normalized performance per area and normalized energy results with respect to the INT16 hardware configuration with the highest performance per area for VGG-16 design space}}\label{fig:vgg}
% \end{figure*}

%% file: conclusion.tex
\vspace{-1mm}
In this work, we present \textit{QAPPA}, a quantization-aware highly parameterized power, performance, and area modeling framework for DNN accelerators. Our framework can foster the future research on design space exploration of DNN accelerators for various design choices such as bit precision, processing element type, scratchpad size of processing elements, global buffer size, device bandwidth, number of total processing elements in the the design, and DNN workloads. 
Our results show that different bit precisions and processing element types lead to significant differences in terms of performance per area and energy. Specifically, LightPE-1 and LightPE-2 achieve $4.9 \times$ and $4.1 \times$ more performance per area and $4.9 \times$ and $4.2 \times$ energy improvement on average when compared to the best INT16 hardware configuration, respectively. Therefore, design space exploration of quantization-aware DNN accelerators merits a meticulous analysis that take these factors into account. 